# Surface-dominant transport in Weyl semimetal NbAs nanowires for next-generation interconnects

Yeryun Cheon[1], Mehrdad T. Kiani[2], Yi-Hsin Tu[3], Sushant Kumar[4], Nghiep Khoan Duong[1], Jiyoung Kim[5], Quynh P. Sam[2], Han Wang[2], Satya K. Kushwaha[6], Nicholas Ng[6], Seng Huat Lee[7], Sam Kielar[5], Chen Li[5], Dimitrios Koumoulis[8], Saif Siddique[2], Zhiqiang Mao[7], Gangtae Jin[9], Zhiting Tian[5], Ravishankar Sundararaman[10], Hsin Lin[11], Gengchiau Liang[3], Ching-Tzu Chen[12], Judy J. Cha[2]*

[1] Department of Physics, Cornell University, Ithaca, New York 14853, United States
[2] Department of Materials Science and Engineering, Cornell University, Ithaca, New York 14853, United States
[3] Industry Academia Innovation School, National Yang-Ming Chiao Tung University, Hsinchu 300093, Taiwan
[4] IBM Research, 257 Fuller Road, Albany, New York 12203, United States
[5] Sibley School of Mechanical and Aerospace Engineering, Cornell University, Ithaca, New York 14853, United States
[6] Platform for the Accelerated Realization, Analysis and Discovery of Interface Materials (PARADIM), The Johns Hopkins University, Baltimore, Maryland 21218, United States
[7] Department of Physics, The Pennsylvania State University, University Park, Pennsylvania 16802, United States
[8] Cornell Center for Materials Research, Cornell University, Ithaca, New York 14853, United States
[9] Department of Electronic Engineering, Gachon University, Seongnam 13120, South Korea
[10] Department of Materials Science and Engineering, Rensselaer Polytechnic Institute, Troy, New York 12180, United States
[11] Institute of Physics, Academia Sinica, Taipei 115201, Taiwan
[12] IBM Thomas J. Watson Research Center, Yorktown Heights, New York 10598, United States

*Corresponding author e-mail: jc476@cornell.edu

**Ongoing demands for smaller and more energy-efficient electronic devices necessitate alternative interconnect materials with lower electrical resistivity at reduced dimensions. Despite the emergence of many promising candidates, synthesizing high-quality nanostructures remains a major bottleneck in evaluating their performance. Here, we report the successful synthesis of Weyl semimetal NbAs nanowires via thermomechanical**

**nanomolding, achieving single crystallinity and controlled diameters as small as 40 nm. Our NbAs nanowires exhibit a remarkably low room-temperature resistivity of 9.7 ± 1.6 μΩ·cm, which is 3–4 times lower than their bulk counterpart. Theoretical calculations corroborate the experimental observations, attributing this exceptional resistivity reduction to surface-dominant conduction with long carrier lifetime at finite temperatures. Further characterization of NbAs nanowires and bulk single crystals reveals high breakdown current density, robust stability, and superior thermal conductivity. Collectively, these properties highlight the strong potential of NbAs nanowires as next-generation interconnects, which can surpass the limitations of current copper-based interconnects. Technologically, our findings present a practical application of topological materials, while scientifically showcasing the fundamental properties uniquely accessible in nanoscale platforms.**

**Main**

With the continued downsizing of electronic components in modern integrated circuits (ICs), current copper (Cu)-based interconnects are reaching their physical limit.[1-4] At the extremely scaled dimensions below the electron mean free path, the electrical resistivity of Cu increases significantly due to the electron scattering at surfaces and grain boundaries. The high resistance of current narrow interconnect lines introduces non-negligible resistance-capacitance (RC) signal delay and elevates dynamic power consumption, limiting the overall computing performance of IC chips.[3,4] As such, addressing these challenges in advanced technology nodes requires alternative conductive materials with lower resistivity than current Cu interconnects at reduced dimensions. While recent studies have focused on elemental metals, such as cobalt (Co) and ruthenium (Ru), which exhibit a less pronounced resistivity scaling effect in the sub-10 nm regime due to their small electron mean free path,[3,4] topological semimetals have emerged as a promising candidate.[5-8] Topological semimetals possess topologically protected conducting surface states, which are predicted to contribute substantially to total conductivity even in large dimensions.[9] At small feature sizes with large surface-to-volume ratio, surface contributions to electrical conduction may dominate over bulk contributions, and certain topological semimetals show, even at room temperature, decreasing resistivity as their dimensions shrink.[10-12] Theoretical studies on Weyl semimetals and multifold-fermion semimetals have indeed demonstrated significant surface-state contributions to electrical conduction at the nanoscale.[5,9,13,14] Despite these encouraging theoretical predictions and a few thin-film studies,[15-17] there remains a noticeable lack of experimental evidence that confirms the superior resistivity scaling of such material systems at dimensions relevant to practical interconnect applications.

Niobium arsenide (NbAs) is a type-I Weyl semimetal that has been extensively studied since its first experimental discovery in 2015.[18] Although a few bottom-up and top-down approaches have been applied to synthesize NbAs nanostructures, they have been limited in terms of achievable sizes or crystal quality. For example, chemical vapor deposition (CVD) growth of NbAs nanobelts with thicknesses of ~200 nm has been reported, and these nanobelts showed an order-of-magnitude reduction in electrical resistivity compared to bulk crystals, demonstrating the high mobility of surface electrons.[11] However, the CVD synthesis lacks fine control over morphology and size, restricting their cross-sectional area to ~$10^5$ nm$^2$, much larger than needed for advanced-node interconnects. On the other hand, NbAs thin films have been grown on gallium arsenide substrates using molecular beam epitaxy, yet they are textured with nanometer-sized grains.[15] As a top-down method, focused ion beam (FIB) milling can produce micro- and nanostructures of NbAs with controlled sizes, but the Ga ion beam leaves Nb residues on their surfaces due to the large difference in surface binding energies between Nb and As.[19]

In this work, we successfully synthesize single-crystal NbAs nanowires with diameters as small as 40 nm using thermomechanical nanomolding (TMNM). TMNM is a recently developed fabrication method through which single-crystal nanowires can be obtained from polycrystalline bulk feedstocks via lattice and interfacial diffusion in porous molds.[20-23] Electrical transport measurements on these NbAs nanowires reveal a desirable resistivity scaling: their room-temperature resistivity (9.7 ± 1.6 μΩ·cm) is 3–4 times lower than that of bulk single crystals and comparable to Cu interconnects in advanced nodes.[24] Magneto-transport measurements at low temperature show clear Shubnikov-de Haas (SdH) oscillations in NbAs nanowires, indicating their high carrier mobility. First-principles calculations and SdH oscillation analysis indicate a negligible chemical potential shift in the NbAs nanowires, ruling out bulk carrier-density variations as the cause of the observed reduction in room-temperature resistivity and thereby suggesting significant contributions from topological surface states. Furthermore, simulations of the electron-phonon resistivity scaling indicate that the surface-state lifetime must exceed the bulk-state lifetime by roughly two orders of magnitude to explain the observed resistivity reduction. Finally, we report the breakdown current density, air sensitivity, and thermal conductivity of NbAs to showcase that NbAs outperforms other candidate materials for future interconnect applications.

**Results**

**Theoretical prediction of electrical transport and resistivity scaling in NbAs**

Fig. 1a shows the crystal structure of NbAs, which belongs to the transition-metal monopnictide (TmPn) family (Tm = Ta/Nb, Pn = P/As) with space group $I4_1md$ (No. 109).[18] This TmPn family lacks inversion symmetry, thereby allowing it to host Weyl fermions. We have performed first-principles calculations of the electronic band structure, phonon band structure, and electron-phonon coupling in NbAs (see **Methods**). Fig. 1b presents the calculated NbAs electronic band structure with finite spin-orbit coupling. NbAs possesses 12 pairs of Weyl nodes, i.e. the discrete linear band crossings near the Fermi level, with topological Fermi-arc surface states connecting the projection of two oppositely charged Weyl nodes in the surface Brillouin zone (BZ). Fig. 1c schematically illustrates the distribution of these Weyl nodes in the three-dimensional BZ and the topologically protected Fermi arcs in the (001) surface-projected two-dimensional BZ.[18,25] Theory predicts the numerous Weyl nodes and surface states yield a comparatively high surface density of states, leading to substantial surface contributions to the total conduction at reduced dimensions.[13]

Fig. 1d shows the calculated bulk carrier density as a function of the chemical potential at 300 K. The red and blue curves represent the electron and hole contributions respectively, highlighting the semimetallic nature of NbAs, i.e., the electron and hole carriers almost perfectly compensate each other at the Fermi level. Fig. 1e shows how the bulk electron-phonon resistivity at 300 K varies with the chemical potential. The elevated resistivity near zero doping primarily stems from the low bulk carrier density. Note that, although carrier doping (e.g., due to vacancies) can reduce the bulk electron-phonon resistivity in NbAs, the chemical potential must shift by ~200 meV for $\gtrsim$ 75% reduction in the resistivity.

Fig. 1f plots the scaling of the room-temperature resistivity in NbAs films and the estimation for NbAs nanowires as a function of the effective surface-to-bulk-state lifetime ratio ($\tau_s/\tau_b$) (see Supplementary Fig. 1 and **Supplementary Note 1** for details). The red, blue, and green traces correspond to $\tau_s/\tau_b$ = 1, 10, and 100, respectively. The open symbols denote the calculated $\rho$ values in 16-atomic-layer (AL) to 56-AL films, in increments of 8 ALs. (One unit cell is 8-AL-thick along the [001] direction in NbAs.) Assuming bulk and surface states conduct in parallel, we fit each calculated $\rho$ dataset (corresponding to a fixed $\tau_s/\tau_b$) using a single fitting parameter $\alpha$ that denotes the ratio of the surface conductance to bulk conductivity[13] (see Supplementary Eq. (5) in **Supplementary Note 1**) and extrapolate the film resistivity to ~40 nm critical dimension (dashed lines in Fig. 1f). We then estimate the corresponding nanowire resistivity scaling by doubling the $\alpha$ values to account for the surface-conductance contribution from the sidewalls besides the top and bottom surfaces (solid lines in Fig. 1f). The resulting sample resistivity reduces with decreasing dimensions because of increasing surface contributions to total conduction. The extent of this reduction increases with the $\tau_s/\tau_b$ ratio, as the higher the $\tau_s/\tau_b$ ratio, the larger the contribution of each surface state. For a ~40 nm diameter NbAs nanowire, a $\gtrsim$ 75% reduction from the bulk resistivity implies $\tau_s/\tau_b \sim 100$ (solid green line in Fig. 1f).

**Synthesis of NbAs nanowires via TMNM**

We first prepared polycrystalline bulk feedstocks of NbAs by spark plasma sintering (see **Methods**). X-ray diffraction confirmed the dominant NbAs phase in the bulk feedstocks, and their electron backscatter diffraction maps showed micron-sized crystal domains (see Supplementary Fig. 2). From these polycrystalline bulk feedstocks, we synthesized single-crystal NbAs nanowires via TMNM. Fig. 2a illustrates a schematic diagram of the TMNM process: a bulk feedstock of the target material is pressed into a porous mold at an elevated temperature. During the TMNM, nanowires grow in specific orientations with energetically favorable surfaces through grain reorientation, allowing the fabrication of single-crystal nanowires even from polycrystalline bulk feedstocks.[23,26] By utilizing molds with different pore sizes, the morphology and size of synthesized nanowires can be controlled. Here, the TMNM was conducted at 700 °C and 200 MPa for 3 hours, using anodic aluminum oxide (AAO) molds with 40 nm diameter pores. Fig. 2b shows a scanning electron microscopy (SEM) image of NbAs nanowires, after the AAO mold was etched (see **Methods**), confirming a uniform wire diameter of ~40 nm. These nanowires were released from the bulk feedstock via sonication in isopropyl alcohol, which results in typical wire lengths of 2–3 μm.

The atomic structure of the NbAs nanowires was characterized using the high-angle annular dark-field scanning transmission electron microscopy (HAADF-STEM). Low- and high-magnification HAADF-STEM images of a NbAs nanowire are shown in Figs. 2c and d, respectively (see Supplementary Fig. 3 for additional NbAs nanowires). These STEM images demonstrate the successful synthesis of single-crystal NbAs nanowires with no noticeable defects, such as stacking faults or dislocations. We also note a thin oxide layer of ~3 nm on the surface (Supplementary Fig. 4a), which is consistent with previous reports on NbAs thin films.[15] Crystallographic analysis using the STEM images and their fast Fourier transform (FFT) reveals that NbAs nanowires grow predominantly along the [100] crystal orientation, although [101] growth orientation was also observed (see Supplementary Figs. 3e and f). For comparison with the nanowires, we also characterized chemical vapor transport grown single crystals. The selected area electron diffraction (SAED) pattern and atomic-resolution HAADF-STEM image of an electron-transparent thin lift-out are shown in Figs. 2e and f, respectively (see Supplementary Fig. 5 for a different crystal orientation). The lattice constants of the nanowires, determined from the FFT of STEM images, are $a$ = 3.48 Å and $c$ = 11.73 Å, which agree with our single crystals ($a$ = 3.45 Å and $c$ = 11.51 Å) and previous studies.[18] Fig. 2g compares the normalized energy-dispersive X-ray spectroscopy (EDX) data of the nanowire and the single crystal. The relative peak intensities of Nb and As indicate that the Nb-to-As ratio is close to 1:1 for the nanowires. At molding temperatures higher than 700 °C, we observed As-deficient phases, such as niobium oxides or metastable $Nb_7As_4$ due to the facile vaporization of arsenic.

**Electrical transport properties of NbAs nanowires and evidence of surface conduction**

To perform electrical transport measurements on NbAs nanowires, we fabricated four-terminal devices using standard e-beam lithography (see **Methods**). The nanowires were dispersed on $SiN_x$/Si substrates and treated with dilute hydrofluoric acid (HF) to remove the thin oxide layer and any residual mold materials. $SiN_x$/Si substrates were used due to the slow etch rate of silicon nitride in HF. Figs. 3a and b show a schematic diagram and an SEM image of a representative NbAs nanowire device, respectively. As shown in Fig. 3c, two-terminal electrical measurements with varied channel lengths reveal linear current-voltage (*I-V*) characteristics at low applied voltages, indicating well-defined ohmic contacts of the devices. Fig. 3d displays the temperature-dependent electrical resistivity of a NbAs nanowire and two bulk single crystals, all of which show metallic behavior. At 300 K, the four-terminal nanowire device shows significantly lower resistivity (~8.1 μΩ·cm) than the bulk (≳ 49 μΩ·cm). Furthermore, the measured nanowire exhibits weak temperature dependence with a low residual-resistivity ratio (RRR) of ~1.8, compared to the single crystals (RRR = ~10 and ~5.6). The surprisingly small electron-phonon resistivity of the nanowire in contrast to that of the bulk single crystals cannot be explained by doping-induced bulk resistivity reduction. A factor of ~4 reduction in resistivity by doping requires varying the chemical potential by ± 200 meV (Fig. 1e), which is unlikely given the near 1:1 stoichiometric ratio of the molded NbAs nanowires (Fig. 2g). As discussed later, quantum oscillation measurements of our samples show that the chemical potential difference between a nanowire and a bulk single crystal is small, further precluding bulk-carrier doping as the cause of the resistivity reduction in our nanowires. This strongly suggests that the low resistivity in NbAs nanowires must contain significant contributions from the surface-state conduction.

Typically, a low RRR suggests poor crystal quality owing to impurities or disorder. However, the residual electrical resistivity of the nanowire remains low and is comparable to that of our bulk single crystals at 2 K (~5 μΩ·cm). Thus, the significantly lower electron-phonon resistivity in NbAs nanowires cannot be explained by poor crystal quality. Instead, it might suggest weak electron-phonon coupling in topological surface states. Indeed, previous studies on topological insulators, such as $Bi_2Se_3$ and $Bi_2Te_3$, have demonstrated weak surface-state electron-phonon coupling using first-principles calculations and angle-resolved photoemission spectroscopy.[27,28]

Fig. 3e summarizes the room-temperature resistivities of molded nanowires, polycrystalline bulk feedstocks, and bulk single crystals of NbAs. Our measured resistivity values for bulk crystals are in good agreement with previously reported values.[11,29,30] At room temperature, which is most relevant for interconnect applications, NbAs nanowires exhibit the average resistivity of 9.7 ± 1.6 μΩ·cm, which is 3–4 times lower than that of best bulk single crystals, and an order of magnitude lower than polycrystalline bulk feedstocks from which the nanowires were grown. The much higher resistivity of polycrystalline bulk feedstocks compared to bulk single crystals is attributed to significant grain boundary scattering, as the mean free path of NbAs, reaching up to a few microns,[29] is comparable to the grain sizes in our bulk feedstocks. Furthermore, comparing the experimental nanowire-to-bulk-resistivity ratio (~0.25 to ~0.33) with the theoretical result (Fig.

1f), we estimate that the average surface-carrier lifetime is roughly two orders of magnitude longer than the bulk-carrier lifetime (see **Supplementary Note 1**), consistent with the weak electron-phonon coupling manifested in the temperature-dependent resistivity data.

We note that the wire diameters used for resistivity calculations were taken either from the nominal pore size of the AAO molds or from low-magnification SEM images. The actual wire diameters are smaller than these nominal values. For example, a previous study on nanomolded $Mo_4P_3$ showed that the actual wire diameter can be 15–20% smaller than the nominal mold diameter.[31] Therefore, the resistivity values reported here represent an upper bound and could be up to ~35% lower in reality.

**Additional materials properties and comparison to current interconnect technologies**

For interconnect applications, low resistivity at small dimensions is not the sole factor for consideration. Other materials properties, such as breakdown current density, thermal conductivity, and resistance to oxidation and electromigration are equally important. Thus, we further evaluated these properties. For the breakdown current density of NbAs nanowires, two-terminal NbAs nanowire devices were fabricated, and the applied voltage was gradually increased until device failure. The measured breakdown current densities were 25.8 $MA/cm^2$, 41.6 $MA/cm^2$, and 78.7 $MA/cm^2$ for three devices, as shown in Fig. 3f. These values are comparable to the breakdown current density of Cu nanowires (~100 $MA/cm^2$),[32,33] and other previously reported nanostructures with high breakdown current density, such as $TaSe_3$ nanowires and $WTe_2$ nanobelts (see Supplementary Table 1).[34-37] The inset of Fig. 3f shows an SEM image of a failed device. Notably, the breakdown spots are not located at the cathode side and appear at random positions; this suggests that device breakdown is likely caused by Joule heating, rather than electromigration-induced void formation.[38,39]

Another important property for interconnect materials is efficient heat dissipation. Here, we measured the thermal conductivity of bulk single crystals and polycrystals of NbAs. Using the laser-induced transient thermal grating (TTG) method, we obtained in-plane thermal diffusivities at room temperature and converted them to thermal conductivities with the measured specific heat of the crystals (see **Methods**). The thermal conductivities of single crystals and polycrystals were determined to be 97.6 ± 3.9 W/m·K and 47.7 ± 4.9 W/m·K, respectively (see Supplementary Fig.

6 and **Supplementary Note 2** for details). As summarized in Fig. 3g, the thermal conductivity of single crystal NbAs is on par with conventional metals such as Ru and Co,[40,41] and significantly higher than or comparable to those of other topological semimetals (see Supplementary Table 2 for details).[42-47] The thermal conductivity of NbAs nanowires was not measured due to the small sample volume relative to the beam size; however, their high electrical conductivity suggests that they may exhibit similar thermal conductivity and heat-transfer capabilities to bulk crystals.

To test air stability, we monitored the electrical resistivity of a nanowire device after exposing it to ambient air (Fig. 3h). The resistivity of the NbAs nanowire increased by ~100 % and plateaued after 1 day of air exposure, suggesting the formation of a self-passivating oxide layer on the nanowire surface. A 40 nm-thick Cu film, sputtered onto a sapphire substrate, was tested as reference under the same conditions. Although it showed better stability than our NbAs nanowire, such thin films expose less surface area than nanowires, thereby mitigating the effects of air exposure. To further investigate the surface oxidation of NbAs, we performed X-ray photoelectron spectroscopy measurements on bulk NbAs crystals annealed in ambient air or argon (Ar) environment at temperatures up to 450 °C. Under Ar environment, we observed negligible surface oxidation up to 200 °C, whereas gradual surface oxidation began above 100 °C in ambient air (see Supplementary Fig. 7 for details). These results indicate the high stability of NbAs in controlled environments at elevated back-end-of-line (BEOL)-compatible temperatures and the need for encapsulation in practical applications. STEM characterization of the surface oxidation in both nanowires and bulk crystals is presented in Supplementary Fig. 4.

Fig. 3i displays the dimensional scaling of room-temperature resistivity of NbAs nanowires and leading alternative metal nanowires for interconnects, including latest Cu (damascene lines with a Co/TaN liner/barrier layer), Ru, and Co.[24,48,49] In contrast to these conventional metals that show increasing resistivity with decreasing dimensions, NbAs exhibits the opposite trend. The NbAs nanowires show resistivity lower than Ru and Co at the same dimension. While the 40 nm diameter NbAs nanowires do not yet surpass the state-of-the-art Cu interconnect technology, the advantage of NbAs is expected to become more pronounced in the sub-10 nm regime as the surface contribution to the electrical conduction exceeds the bulk contribution.[13] This makes NbAs a promising candidate for next-generation interconnects at 16 nm pitch and below.

**Quantum oscillations of NbAs nanowires and chemical potential estimation**

We performed magneto-transport measurements on both a bulk single crystal and a nanowire of NbAs at 2 K and compared their perpendicular magnetoresistance (MR), as shown in Figs. 4a and b. At 2 K, the bulk crystal exhibits a large MR of up to ~13,000% at 9 T with the magnetic field applied along the crystallographic *c*-axis. In contrast, the nanowire shows a much lower MR of up to ~2.5% at 9 T. This behavior is commonly observed in nanowires and thin films due to the confinement and scattering effects.[50,51] Despite this lower MR, the NbAs nanowire displays clear Shubnikov-de Haas (SdH) oscillations similar to those observed in bulk crystals. The oscillatory component of the resistivity is obtained by subtracting a polynomial background and is plotted as a function of $1/B$ in Fig. 4c. From the FFT of these background-subtracted resistivities, two dominant oscillation frequencies are identified for each sample: 13.8 T and 24.2 T for the bulk crystal, and 8.8 T and 29.4 T for the nanowire (Figs. 4d and e). These SdH oscillation frequencies are consistent with previous reports on NbAs bulk crystals.[29,52,53] Note that the frequency range corresponding to the Fermi-arc states is at least an order of magnitude higher than our experimental data.[11] Thus, the SdH oscillations we have measured all originate from the bulk Fermi surfaces.

Fig. 4f shows the calculated SdH oscillation frequencies as a function of the chemical potential with the magnetic field oriented along the *c*-axis (see **Supplementary Note 3**). These frequencies are derived from the extremal cross-section areas of electron and hole pockets in the computed bulk band structures. We have also calculated the SdH oscillation frequencies as a function of the tilting angle of the magnetic field from the *c*-axis (see Supplementary Fig. 8f), which shows the SdH frequencies increase with the tilting angle in all directions.

Compared to the experimental data, we see that the chemical potential of the bulk single crystal lies 7–9 meV below zero doping, while that of the nanowire device lies 14–18 meV below (assuming magnetic fields parallel to the *c*-axis), implying these samples are hole-doped. The spread in the estimated hole- and electron-chemical potential values can be attributed to the misalignment between the magnetic field and the *c*-axis of the samples. See **Supplementary Note 3** for more detailed explanations.

Based on the above analysis, the chemical potential difference between the nanowire and the bulk crystal is small, even considering the uncertainty in the field orientation. Our earlier analysis (Fig. 1e) shows that a Fermi level shift of ~200 meV would be required to explain a three-fold resistivity reduction solely via bulk conduction. Furthermore, all three measured NbAs nanowires exhibit much reduced room-temperature resistivity compared to the bulk, even though the Fermi level for each may likely vary. Combining the quantum oscillation analyses above, we thus prove that bulk resistivity variation alone leads to a negligible resistivity reduction in our nanowire samples. Hence, the low room-temperature resistivity of our nanowires must predominantly come from the Fermi-arc surface states with a significantly longer scattering lifetime than the bulk states (Fig. 1f).

We note that the SdH oscillations remain visible in the nanowire device after 2 months of air exposure, which highlights the exceptional stability of NbAs at the nanoscale. Since these quantum oscillations originate from the coherent motion of quasi-particles near the Fermi surface and are therefore highly sensitive to defects, their persistence further verifies the preserved crystalline quality of our NbAs nanowires.

**Benchmark comparison of NbAs among promising topological semimetals for interconnects**

Fig. 5a compares the room-temperature electrical resistivities of NbAs nanowires with those of other promising topological semimetals in nanowire geometry (CoSi, MoP, WP, $Cd_3As_2$, $MoP_2$, and $TaAs_2$).[10,43,54-58] At sub-100 nm dimensions, NbAs nanowires exhibit the lowest room-temperature resistivity among the reported topological semimetal nanowires. To gauge the surface-state contributions to electrical conduction, we also compared the nanowire resistivity scaled by the respective bulk resistivity, defined as $\rho(nanowire)/\rho(bulk)$, as shown in Fig. 5b. It is evident that NbAs and the multifold-fermion semimetal cobalt monosilicide (CoSi) nanowires exhibit surface-dominant electrical conduction at reduced dimensions. In Figs. 5a and b, we focus only on experimental results from nanowires, as direct comparison between nanowires and nanostructures in different geometries would not be proper. In general, thin films tend to exhibit lower resistivity than 1D-confined nanowires, even at similar crystal quality, and our work targets extremely scaled interconnect lines whose morphology takes on the nanowire geometry.

In Figs. 5c and d, we summarize the room-temperature resistivity scaling in topological semimetal thin films (NbAs, CoSi, NbP, TaAs, and $Co_3Sn_2S_2$)[12,15,16,59,60] and thin flakes/nanobelts (NbAs, $WTe_2$, $ZrTe_5$, $PtBi_2$, and $SnTaS_2$).[11,61-64] Among the presented nanostructures, thin films of amorphous NbP on epitaxial Nb seed layers and thin flakes of $SnTaS_2$ show a desirable trend of decreasing resistivity with reducing thickness.[12,64] Remarkably, even when compared with various nanostructures of different geometries, NbAs demonstrates the most promising results.

We note that the resistivity of our NbAs nanowires is higher than the previously reported value of ~3 μΩ·cm for CVD-grown NbAs nanobelts at ~200 nm thickness (see Figs. 5c and d).[11] This discrepancy may arise from differences in sample geometry (nanowires vs. nanobelts) and surface roughness. Compared to CVD-grown nanobelts, nanowires have a smaller fraction of the (001) surface that hosts the topologically protected Fermi arcs. Although other surfaces such as (100) can also host conducting surface states, they are not guaranteed by nontrivial topology and may disappear with doping.

For practical interconnect applications, we further compared the line resistance of NbAs nanowires with industry projections for Cu and Ru damascene lines assuming the current liner/barrier layers (see Supplementary Fig. 9).[24] NbAs nanowires with cross-section areas of 1250–1400 nm$^2$ are expected to exhibit similar line resistance to Ru lines but higher than Cu lines. Future investigations with smaller sample dimensions are needed to explore more advanced downscaling behavior.

**Conclusions**

Nearly two decades have passed since the first experimental discovery of topological materials, which has yielded numerous groundbreaking insights into their fundamental properties. Yet, realistic applications of these materials remain elusive. As demonstrated here with Weyl semimetal NbAs nanowires, low-resistance next-generation interconnects may be among the first realizations of such applications. To fully capitalize on this potential, however, further investigations are warranted, including more in-depth transport studies with varied wire diameters and refined theoretical frameworks that account for electron-phonon interactions in topological surface states within nanowire geometries. Nevertheless, our results mark a critical step forward in bridging the gap between fundamental discoveries and practical applications, ultimately realizing the long-anticipated promise of topological semimetals.

**Methods**
**Bulk feedstock preparation**

Niobium arsenide powder with an atomic ratio of 1:1 and a purity of 99.99% was purchased from American Elements. As-received powder was sintered via spark plasma sintering using a LABOX-650F in a 10 mm diameter graphite die at 1000 °C and 100 MPa for 10 minutes under vacuum. The sintered samples were then vacuum sealed in a quartz tube and annealed at 1000 °C for 2 days. The annealed samples are then cut and polished to mirror finish before molding.

**Thermomechanical nanomolding (TMNM)**

A mirror-polished bulk feedstock was placed on a porous AAO membrane (InRedox, 40 nm pore size, 50 μm thickness) in a graphite die. The sample was heated to 700 °C with a ramp rate of 10 °C/min and pressed at 200 MPa for 3 hours in an argon environment ($H_2O < 0.5$ ppm, $O_2 < 0.5$ ppm). After the duration, the pressed sample was cooled naturally. The AAO mold was then etched away using 48% concentrated HF at room temperature for 1-2 days. As-grown nanowires were separated from the bulk feedstock by sonicating in isopropyl alcohol (IPA).

**Single crystal growth**

The NbAs single crystals were provided by the Penn State 2D Crystal Consortium-Materials Innovation Platform (2DCC-MIP). The single crystals were grown using the chemical vapor transport (CVT) method with iodine as the transport agent. Initially, a stoichiometric mixture of Nb powder and As chunk was heated to 800 °C for 72 hours in an evacuated quartz tube to form polycrystalline NbAs. This polycrystalline NbAs was then transferred into a new evacuated quartz tube with 10 mg/cm$^3$ of iodine. The sealed ampoule was placed in a four-zone tube furnace and heated to 950 °C and 850 °C at the charges zone and growth zone, respectively, for 4 days. This resulted in the formation of multiple NbAs crystals with distinct well-faceted, flat plate-like morphology.

**Materials characterization**

X-ray diffraction measurements on polycrystalline bulk feedstocks and single crystals were carried out using a Bruker D8 Advance ECO powder diffractometer with a Cu Kα X-ray source. Scanning electron microscopy (SEM) images were taken on a Zeiss Sigma 500 Scanning Electron Microscope. Electron backscatter diffraction (EBSD) maps were taken on polished feedstock samples using a Bruker eFlash HR EBSD on a Tescan Mira3 FESEM. X-ray photoelectron spectroscopy characterization was performed on bulk crystals using a Thermo Scientific Nexsa G2 Surface Analysis System.

Nanomolded wires were drop cast onto TEM grids. Scanning transmission electron microscopy (STEM) images were taken using a Thermo Fisher Spectra 300 STEM at 300 kV. Lift-outs of single crystals were made using a Helios G4 UX Focused Ion Beam. The final milling was performed at 2 kV to minimize surface damage.

**Device fabrication and electrical measurements**

As-molded NbAs nanowires were drop cast on a $SiN_x$/Si substrate and treated with diluted HF for 2 minutes. The sample was then spin-coated with a double layer of e-beam resist (PMMA 495K A4, 200 nm each). Four- and two-terminal electrode patterns were made by standard e-beam lithography using a Zeiss Supra 55 SEM equipped with Nabity Nanometer Pattern Generator System (NPGS). Ex-situ Ar etching (50 W, 30 sccm, 50 mTorr, 2 minutes) was followed by 10/100 nm-thick Cr/Au deposition using an e-beam evaporator (CVC SC4500 evaporation system). Fabricated devices were annealed at 150 °C for 2 hours under high vacuum (~$10^{-6}$ mbar) to improve the contact.

Room-temperature electrical measurements were carried out in a Lakeshore CRX-VF cryogenic probe station under high vacuum (~$10^{-6}$ mbar), using an Agilent B1500A semiconductor device analyzer or a low-frequency lock-in amplifier (Stanford Research SR830) with a 1 MΩ resistor. Low-temperature electrical measurements were performed using a Quantum Design Physical Property Measurement System (PPMS) down to the base temperature of 2 K. For magneto-transport measurements, perpendicular magnetic fields were swept from -9 T to 9 T at a rate of 80 Oe/s.

**Thermal conductivity measurements**

The in-plane thermal conductivity of bulk single crystals and polycrystals of NbAs was measured using the laser-induced transient thermal grating (TTG) method with a home-built experimental setup.[65-67] Before the measurements, the samples were mirror-polished to remove surface oxide layers and to achieve the nanometer-scale surface roughness. In the TTG system, two pulsed pump beams are focused onto the sample surface, creating an optical interference pattern that leads to a thermal grating. The decay signal of the thermal grating is detected by the diffracted continuous-wave probe beams. By analyzing the decay rate, the in-plane thermal diffusivity is determined and subsequently converted to the thermal conductivity using the density and specific heat capacity of the samples. Further details of the measurements are provided in Supplementary Note 2.

## Theoretical calculations

We construct the Wannier tight-binding model of bulk NbAs with a basis set of maximally-localized Wannier functions[68] using the open-source plane-wave DFT code JDFTx.[69] We employ fully relativistic optimized norm-conserving Vanderbilt pseudopotentials (ONCVPSP)[70] as distributed by the open-source PSEUDODOJO library.[70] These DFT calculations are carried out using the Perdew-Burke-Ernzerhof (PBE) generalized gradient approximation (GGA) for the exchange-correlation functional,[71] with cutoffs of 40 Hartrees and 200 Hartrees for plane-wave and charge density respectively. The pseudopotentials self-consistently include the spin-orbit coupling. We use a $k$-point sampling of $8 \times 8 \times 8$ and Fermi smearing of 0.01 Hartrees for the DFT calculations. For phonon calculations, we use a $q$-point mesh of $2 \times 2 \times 2$. The Wannier functions are constructed using the $p$-orbitals of As and $d$-orbitals of Nb as shown in the previous work.[13] The electrical resistivity at any chemical potential is evaluated using the linearized Boltzmann equation with a full-band relaxation time approximation. To obtain converged integrals over the Brillouin zone, we perform the Wannier interpolation of the electronic energies, phonon frequencies, and electron-phonon matrix elements to a much finer $k$- and $q$-mesh. The details of this methodology used can be found in the other work.[72] Further calculation details are provided in Supplementary Notes 1 and 3.

## Data availability

The data that support the findings of this study are available from the corresponding author upon reasonable request.

## Acknowledgements

Nanowire synthesis was in part supported by the Gordon and Betty Moore Foundation's EPiQS Initiative, Grant GBMF9062.01. Electrical transport measurements were supported by the Semiconductor Research Corporation JUMP 2.0 SUPREME. S.K. and R.S. acknowledge funding from Semiconductor Research Corporation under Task no. 2966.002. H.L. acknowledges the support by Academia Sinica in Taiwan under grant number AS-iMATE-113-15. First-principles electron transport calculations were performed at the Center for Computational Innovations at Rensselaer Polytechnic Institute. Y.C. was supported by the ILJU Graduate Fellowship. Q.P.S. was supported by the National Science Foundation (NSF) GRFP under Grant No. 2139899. This work made use of the Cornell Center for Materials Research shared instrumentation facility. Thermo-Fisher Helios G4 UX FIB and Kraken STEM acquisition was supported by the NSF (DMR-2039380). Device fabrication was performed in part at the Cornell NanoScale Facility, a member of the National Nanotechnology Coordinated Infrastructure (NNCI), which is supported by the NSF (Grant NNCI-2025233). Thermal transport measurements were supported by the SRC JUMP 2.0 SUPREME Seed Grant, the SRC JUMP 2.0 CHIMES, and DARPA YFA (D23AP00159-00). Support for single crystal growth was provided by the NSF through the Penn State 2D Crystal Consortium-Materials Innovation Platform (2DCC-MIP) under NSF cooperative agreement DMR-2039351. Bulk feedstock preparation made use of the synthesis facility of the Platform for the Accelerated Realization, Analysis, and Discovery of Interface Materials (PARADIM), which is supported by the NSF under Cooperative Agreement No. DMR-2039380. G.L. and Y.-H.T. are supported by the National Science and Technology Council (NSTC) under grant number NSTC 112-2112-M-A49 -047 -MY3.

## Author information

**Contributions.** Y.C. synthesized the NbAs nanowires, fabricated the devices, and performed electrical and magneto-transport measurements. Y.-H.T., S.K., R.S., H.L., G.L., and C.-T.C. conducted theoretical analyses. M.T.K. prepared lift-outs of the single crystals for STEM characterization. M.T.K., N.K.D., Q.P.S., and S.S. carried out STEM characterization. J.K., S.K., C.L., and Z.T. performed thermal conductivity measurements and modeling. M.T.K., G.J., S.K.K., and N.N. synthesized the bulk feedstocks. S.H.L. and Z.M. provided the NbAs single crystals. H.W. and G.J. assisted with device fabrication and electrical transport measurements. Y.C. and D.K. performed heat capacity measurements. J.J.C. supervised the project. Y.C., Y.-H.T., S.K., C.-T.C., and J.J.C. wrote the manuscript with input from all authors. All authors reviewed and approved the final version of the manuscript.

**Corresponding author.** Correspondence to Judy J. Cha

## Ethics declarations

## Supplementary information

Supplementary Notes 1–3, Supplementary Figs. 1–9, Supplementary Tables 1 and 2.

## References

1  Gall, D. *et al.* Materials for interconnects. *MRS Bull.* **46**, 959-966 (2021).
2  Moon, J. H. *et al.* Materials Quest for Advanced Interconnect Metallization in Integrated Circuits. *Adv. Sci.* **10**, 2207321 (2023).
3  Kim, J.-S. *et al.* Addressing interconnect challenges for enhanced computing performance. *Science* **386**, eadk6189 (2024).
4  Soulié, J.-P. *et al.* Selecting alternative metals for advanced interconnects. *J. Appl. Phys.* **136**, 171101 (2024).
5  Chen, C.-T. *et al.* Topological Semimetals for Scaled Back-End-Of-Line Interconnect Beyond Cu. *2020 IEEE International Electron Devices Meeting (IEDM)*, pp. 32.4.1-32.4.4 (2020).
6  Han, H. J., Liu, P. & Cha, J. J. 1D topological systems for next-generation electronics. *Matter* **4**, 2596-2598 (2021).
7  Zhai, E. *et al.* The rise of semi-metal electronics. *Nat. Rev. Electr. Eng.* **1**, 497-515 (2024).
8  Kim, S. H. *et al.* Topological semimetals for advanced node interconnects. *iScience* **27**, 111460 (2024).
9  Breitkreiz, M. & Brouwer, P. W. Large Contribution of Fermi Arcs to the Conductivity of Topological Metals. *Phys. Rev. Lett.* **123**, 066804 (2019).
10  Tsai, C.-I. *et al.* Cobalt Silicide Nanostructures: Synthesis, Electron Transport, and Field Emission Properties. *Cryst. Growth Des.* **9**, 4514-4518 (2009).
11  Zhang, C. *et al.* Ultrahigh conductivity in Weyl semimetal NbAs nanobelts. *Nat. Mater.* **18**, 482-488 (2019).
12  Khan, A. I. *et al.* Surface conduction and reduced electrical resistivity in ultrathin noncrystalline NbP semimetal. *Science* **387**, 62-67 (2025).
13  Kumar, S. *et al.* Surface-dominated conductance scaling in Weyl semimetal NbAs. *npj Comput. Mater.* **10**, 84 (2024).
14  Lien, S.-W. *et al.* Unconventional resistivity scaling in topological semimetal CoSi. *npj Quantum Mater.* **8**, 3 (2023).
15  Yánez-Parreño, W. *et al.* Thin film growth of the Weyl semimetal NbAs. *Phys. Rev. Mater.* **8**, 034204 (2024).
16  Nelson, J. N. *et al.* Thin-film TaAs: Developing a platform for Weyl semimetal devices. *Matter* **6**, 2886-2899 (2023).
17  Bedoya-Pinto, A. *et al.* Realization of Epitaxial NbP and TaP Weyl Semimetal Thin Films. *ACS Nano* **14**, 4405-4413 (2020).
18  Xu, S.-Y. *et al.* Discovery of a Weyl fermion state with Fermi arcs in niobium arsenide. *Nat. Phys.* **11**, 748-754 (2015).

19   Bachmann, M. D. *et al.* Inducing superconductivity in Weyl semimetal microstructures by selective ion sputtering. *Sci. Adv.* **3**, e1602983 (2017).
20   Kiani, M. T. & Cha, J. J. Nanomolding of topological nanowires. *APL Mater.* **10**, 080904 (2022).
21   Liu, N. *et al.* General Nanomolding of Ordered Phases. *Phys. Rev. Lett.* **124**, 036102 (2020).
22   Liu, Z., Han, G., Sohn, S., Liu, N. & Schroers, J. Nanomolding of Crystalline Metals: The Smaller the Easier. *Phys. Rev. Lett.* **122**, 036101 (2019).
23   Liu, N. *et al.* Unleashing nanofabrication through thermomechanical nanomolding. *Sci. Adv.* **7**, eabi4567 (2021).
24   Edelstein, D. *et al.* The Extreme Extendibility of Cu and Post-Cu Dual Damascene BEOL Interconnect Technology. *2024 IEEE International Electron Devices Meeting (IEDM)*, pp. 28.3.1-28.3.4 (2024).
25   Lee, C.-C. *et al.* Fermi surface interconnectivity and topology in Weyl fermion semimetals TaAs, TaP, NbAs, and NbP. *Phys. Rev. B* **92**, 235104 (2015).
26   Sam, Q. P. *et al.* Nanomolding of Two-Dimensional Materials. *ACS Nano* **18**, 1110-1117 (2024).
27   Heid, R., Sklyadneva, I. Y. & Chulkov, E. V. Electron-phonon coupling in topological surface states: The role of polar optical modes. *Sci. Rep.* **7**, 1095 (2017).
28   Pan, Z.-H. *et al.* Measurement of an exceptionally weak electron-phonon coupling on the surface of the topological insulator $Bi_2Se_3$ using angle-resolved photoemission spectroscopy. *Phys. Rev. Lett.* **108**, 187001 (2012).
29   Luo, Y. *et al.* Electron-hole compensation effect between topologically trivial electrons and nontrivial holes in NbAs. *Phys. Rev. B* **92**, 205134 (2015).
30   Ghimire, N. J. *et al.* Magnetotransport of single crystalline NbAs. *J. Phys.: Condens. Matter* **27**, 152201 (2015).
31   Kiani, M. T. *et al.* Nanomolding of metastable $Mo_4P_3$. *Matter* **6**, 1894-1902 (2023).
32   Li, L., Zhu, Z., Yoon, A. & Wong, H. S. P. In-Situ Grown Graphene Enabled Copper Interconnects With Improved Electromigration Reliability. *IEEE Electron Device Lett.* **40**, 815-817 (2019).
33   Huang, Q., Lilley, C. M., Bode, M. & Divan, R. Surface and size effects on the electrical properties of Cu nanowires. *J. Appl. Phys.* **104**, 023709 (2008).
34   Stolyarov, M. A. *et al.* Breakdown current density in h-BN-capped quasi-1D $TaSe_3$ metallic nanowires: prospects of interconnect applications. *Nanoscale* **8**, 15774-15782 (2016).
35   Empante, T. A. *et al.* Low Resistivity and High Breakdown Current Density of 10 nm Diameter van der Waals $TaSe_3$ Nanowires by Chemical Vapor Deposition. *Nano Lett.* **19**, 4355-4361 (2019).
36   Mleczko, M. J. *et al.* High Current Density and Low Thermal Conductivity of Atomically Thin Semimetallic $WTe_2$. *ACS Nano* **10**, 7507-7514 (2016).
37   Song, S. *et al.* Electrically Robust Single-Crystalline $WTe_2$ Nanobelts for Nanoscale Electrical Interconnects. *Adv. Sci.* **6**, 1801370 (2019).
38   Tu, K. N. Recent advances on electromigration in very-large-scale-integration of interconnects. *J. Appl. Phys.* **94**, 5451-5473 (2003).
39   Misra, E., Marenco, C., Theodore, N. D. & Alford, T. L. Failure mechanisms of silver and aluminum on titanium nitride under high current stress. *Thin Solid Films* **474**, 235-244 (2005).

40  Ho, C. Y., Powell, R. W. & Liley, P. E. Thermal Conductivity of the Elements. *J. Phys. Chem. Ref. Data* **1**, 279-421 (1972).
41  Cancellieri, C. *et al.* Interface and layer periodicity effects on the thermal conductivity of copper-based nanomultilayers with tungsten, tantalum, and tantalum nitride diffusion barriers. *J. Appl. Phys.* **128**, 195302 (2020).
42  Watzman, S. J. *et al.* Dirac dispersion generates unusually large Nernst effect in Weyl semimetals. *Phys. Rev. B* **97**, 161404(R) (2018).
43  Han, H. J. *et al.* Topological Metal MoP Nanowire for Interconnect. *Adv. Mater.* **35**, 2208965 (2023).
44  Xiang, J. *et al.* Anisotropic thermal and electrical transport of Weyl semimetal TaAs. *J. Phys.: Condens. Matter* **29**, 485501 (2017).
45  Qian, X. *et al.* Anisotropic thermal transport in van der Waals layered alloys $WSe_{2(1-x)}Te_{2x}$. *Appl. Phys. Lett.* **112**, 241901 (2018).
46  Sk, S., Shahi, N. & Pandey, S. K. Experimental and computational approaches to study the high temperature thermoelectric properties of novel topological semimetal CoSi. *J. Phys.: Condens. Matter* **34**, 265901 (2022).
47  Zhang, C. *et al.* Unexpected low thermal conductivity and large power factor in Dirac semimetal $Cd_3As_2$. *Chin. Phys. B* **25**, 017202 (2016).
48  Wen, L. G. *et al.* Ruthenium Metallization for Advanced Interconnects. *2016 IEEE International Interconnect Technology Conference / Advanced Metallization Conference (IITC/AMC)*, 34-36 (2016).
49  Yoo, E. *et al.* Electrical resistivity and microstructural evolution of electrodeposited Co and Co-W nanowires. *Mater. Charact.* **166**, 110451 (2020).
50  Heremans, J., Thrush, C. M., Lin, Y.-M., Cronin, S. B. & Dresselhaus, M. S. Transport properties of antimony nanowires. *Phys. Rev. B* **63**, 085406 (2001).
51  Oyarzún, S. *et al.* Transverse magnetoresistance induced by electron-surface scattering on thin gold films: Experiment and theory. *Appl. Surf. Sci.* **289**, 167-172 (2014).
52  Komada, M. *et al.* Angle-dependent nontrivial phase in the Weyl semimetal NbAs with anisotropic Fermi surface. *Phys. Rev. B* **101**, 045135 (2020).
53  Naumann, M. *et al.* Weyl Nodes Close to the Fermi Energy in NbAs. *Phys. Status Solidi B* **259**, 2100165 (2021).
54  Jin, G. *et al.* Vapor phase synthesis of topological semimetal $MoP_2$ nanowires and their resistivity. *Appl. Phys. Lett.* **121**, 113105 (2022).
55  Jin, G. *et al.* Diameter-dependent phase selectivity in 1D-confined tungsten phosphides. *Nat. Commun.* **15**, 5889 (2024).
56  Roy, A. *et al.* Intertwined topological phases in $TaAs_2$ nanowires with giant magnetoresistance and quantum coherent surface transport. *arXiv:2411.15974* (2024).
57  Li, C. Z. *et al.* Giant negative magnetoresistance induced by the chiral anomaly in individual $Cd_3As_2$ nanowires. *Nat. Commun.* **6**, 10137 (2015).
58  Wang, L. X. *et al.* Magnetotransport properties near the Dirac point of Dirac semimetal $Cd_3As_2$ nanowires. *J. Phys.: Condens. Matter* **29**, 044003 (2017).
59  Rocchino, L. *et al.* Unconventional magnetoresistance and resistivity scaling in amorphous CoSi thin films. *Sci. Rep.* **14**, 20608 (2024).
60  Ikeda, J. *et al.* Critical thickness for the emergence of Weyl features in $Co_3Sn_2S_2$ thin films. *Commun. Mater.* **2**, 18 (2021).

61	Na, J. *et al.* Tuning the magnetoresistance of ultrathin $WTe_2$ sheets by electrostatic gating. *Nanoscale* **8**, 18703-18709 (2016).
62	Niu, J. *et al.* Electrical transport in nanothick $ZrTe_5$ sheets: From three to two dimensions. *Phys. Rev. B* **95**, 035420 (2017).
63	Zhu, A. *et al.* Thickness-tuned magnetotransport properties of topological semimetal trigonal $PtBi_2$. *Appl. Phys. Lett.* **122**, 113101 (2023).
64	Gao, W. *et al.* Evidences of Topological Surface States in the Nodal-Line Semimetal $SnTaS_2$ Nanoflakes. *ACS Nano* **17**, 4913-4921 (2023).
65	Li, C., Ma, Y. & Tian, Z. Thermal Switching of Thermoresponsive Polymer Aqueous Solutions. *ACS Macro Lett.* **7**, 53-58 (2018).
66	Li, C. *et al.* Remarkably Weak Anisotropy in Thermal Conductivity of Two-Dimensional Hybrid Perovskite Butylammonium Lead Iodide Crystals. *Nano Lett.* **21**, 3708-3714 (2021).
67	Chang, B. S. *et al.* Thermal Percolation in Well-Defined Nanocomposite Thin Films. *ACS Appl. Mater. Interfaces* **14**, 14579-14587 (2022).
68	Marzari, N., Mostofi, A. A., Yates, J. R., Souza, I. & Vanderbilt, D. Maximally localized Wannier functions: Theory and applications. *Rev. Mod. Phys.* **84**, 1419-1475 (2012).
69	Sundararaman, R. *et al.* JDFTx: software for joint density-functional theory. *SoftwareX* **6**, 278-284 (2017).
70	van Setten, M. J. *et al.* The PseudoDojo: Training and grading a 85 element optimized norm-conserving pseudopotential table. *Comput. Phys. Commun.* **226**, 39-54 (2018).
71	Perdew, J. P., Burke, K. & Ernzerhof, M. Generalized Gradient Approximation Made Simple. *Phys. Rev. Lett.* **77**, 3865 (1996).
72	Kumar, S., Multunas, C. & Sundararaman, R. Fermi surface anisotropy in plasmonic metals increases the potential for efficient hot carrier extraction. *Phys. Rev. Mater.* **6**, 125201 (2022).

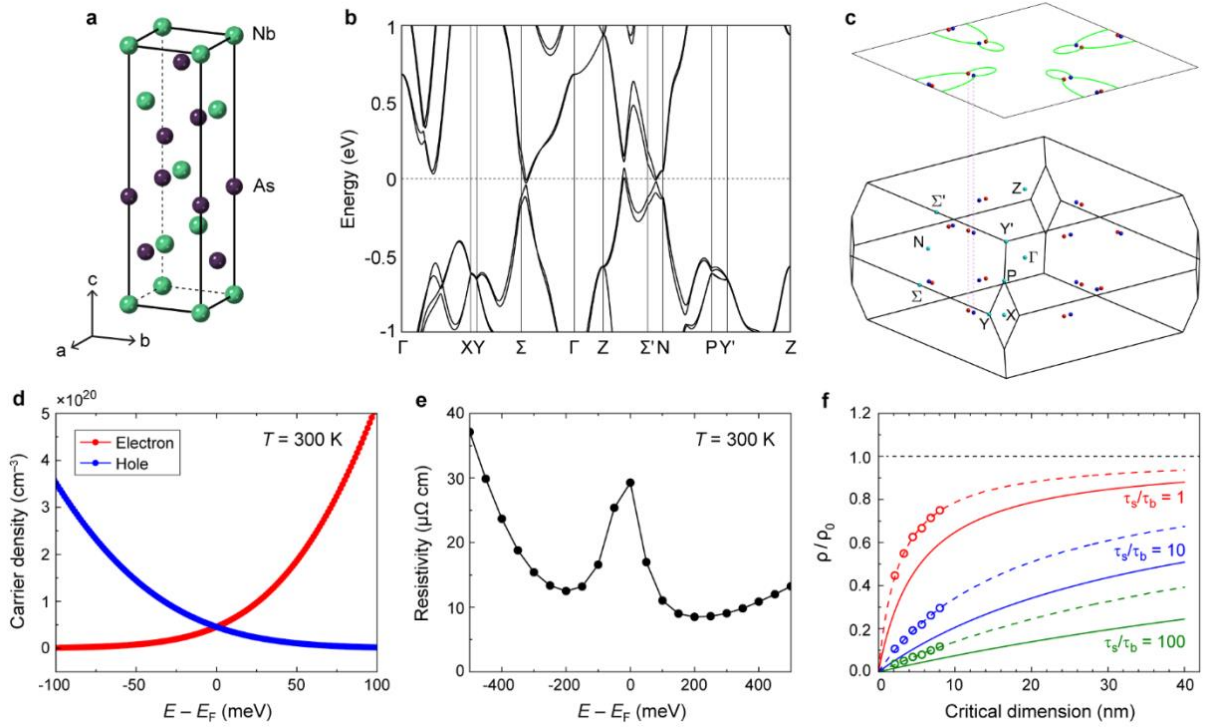

**Fig. 1. Theoretical calculations on Weyl semimetal NbAs. a,** Crystal structure of NbAs. **b,** Calculated electronic band structure of NbAs with finite spin-orbit coupling. **c,** Schematic illustration of the Weyl nodes in the three-dimensional Brillouin zone (BZ) of NbAs and their projections onto the (001) surface BZ. Red and blue denote the opposite chirality of the nodes. Fermi arcs (light green) connect the projected Weyl nodes that are oppositely charged. **d,** Electron (red) and hole (blue) carrier densities at 300 K as a function of the chemical potential. **e,** Calculated bulk electron-phonon resistivity vs. chemical potential at 300 K. **f,** Simulated room-temperature resistivity $\rho/\rho_0$ (where $\rho_0$ is the bulk resistivity) vs. critical dimension (CD) as a function of the surface-to-bulk-state lifetime ratio, $\tau_s/\tau_b = 1$, 10, and 100 (red, blue, and green traces, respectively). $\tau_s$ and $\tau_b$ denote the surface- and bulk-state lifetime, respectively. For thin films, CD refers to the thickness; for nanowires, CD refers to the diameter. The open symbols are the calculated $\rho$ values in 16-AL to 56-AL films, in increments of 8 ALs. The dash lines are obtained by fitting Supplementary Eq. (5) to each dataset for thin films, and the solid lines are the approximated resistivity scaling for nanowires.

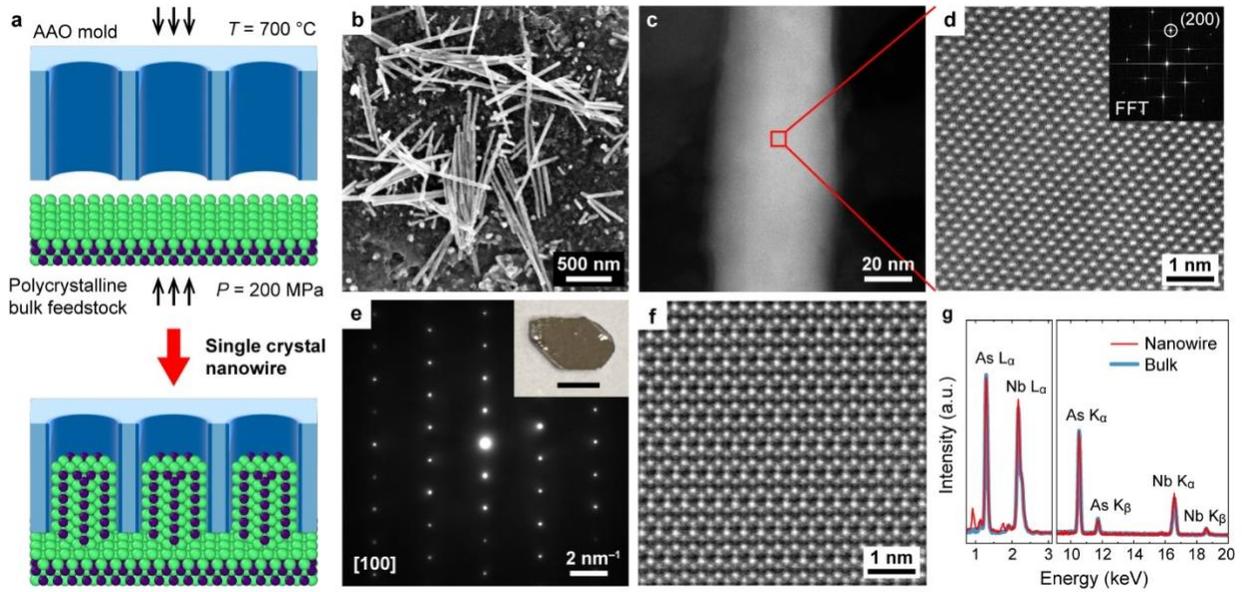

**Fig. 2. Thermomechanical nanomolding (TMNM) and structural characterization of NbAs nanowires. a,** Schematic illustration of the TMNM process. **b,** SEM image of NbAs nanowires still attached to a bulk feedstock. **c,** Low-magnification HAADF-STEM image of a NbAs nanowire. **d,** Atomic-resolution HAADF-STEM image of the region outlined by the red square in (**c**). The inset shows the corresponding FFT with the [100] growth orientation indicated. **e,** SAED pattern of a lift-out from a bulk single crystal of NbAs. The inset shows an optical photograph of the single crystal (black scale bar: 1 mm). **f,** Atomic-resolution HAADF-STEM image of the same lift-out shown in (**e**). **g,** Normalized STEM-EDX spectra of the nanowire (red) and the bulk single crystal (blue).

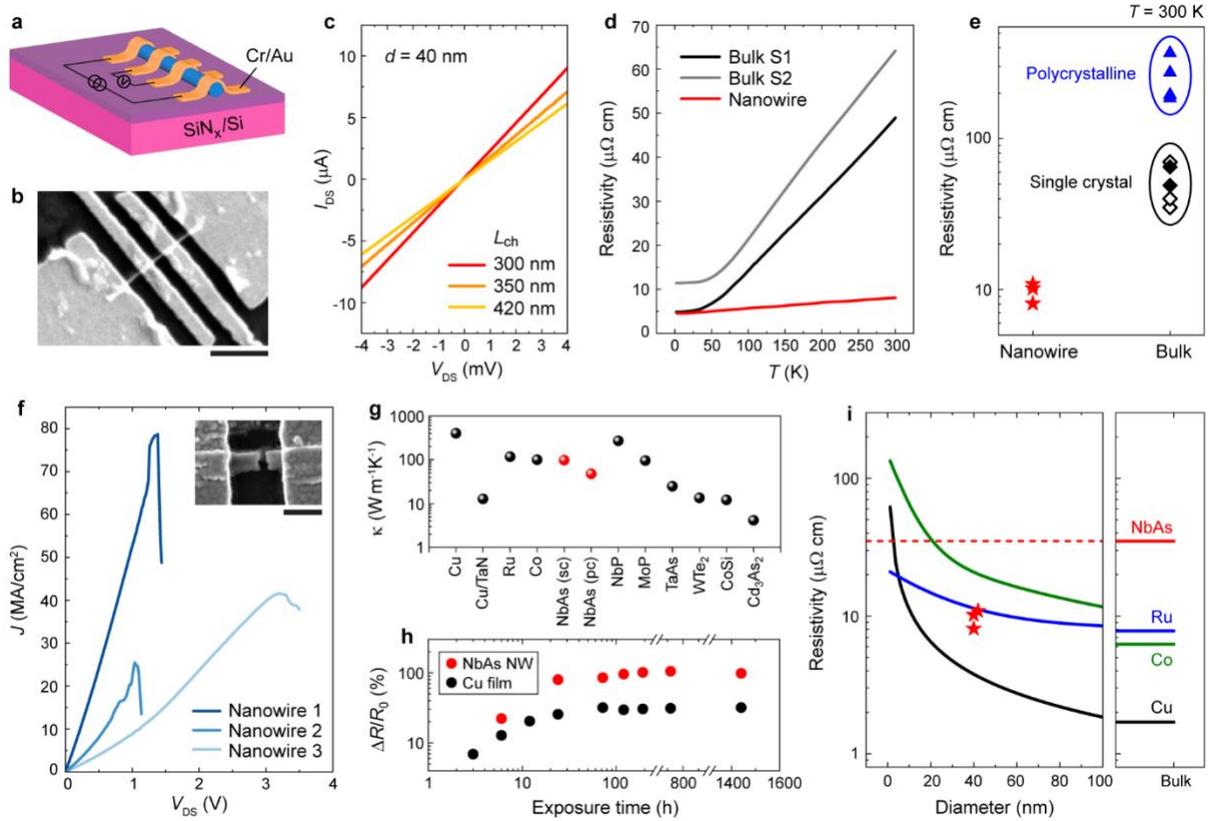

**Fig. 3. Electrical properties of NbAs nanowires and additional materials characterizations for interconnect applications. a,** Schematic diagram of a four-terminal NbAs nanowire device. **b,** Representative SEM image of a four-terminal NbAs nanowire device (scale bar: 1 μm). **c,** Two-terminal *I-V* curves for different channel lengths ($L_{ch}$). **d,** Temperature-dependent four-terminal electrical resistivity for bulk single crystals (black and gray) and a nanowire (red). **e,** Room-temperature resistivities of polycrystalline bulk feedstocks (blue triangles), bulk single crystals (black rhombuses; solid symbols from this work and open symbols from literature[11,29,30]), and nanowires (red stars). **f,** Current density versus applied voltage for three different two-terminal nanowire devices. The inset shows a representative SEM image of a failed device (scale bar: 100 nm). **g,** Thermal conductivity of various materials, including NbAs from this work (where sc and pc denote single crystals and polycrystals, respectively), conventional metals (Cu, Ru, and Co),[40,41] and other representative topological semimetals (NbP, MoP, TaAs, WTe$_2$, CoSi, and Cd$_3$As$_2$).[42-47] **h,** Normalized resistance changes ($\Delta R/R_0$) versus air exposure time for a 40 nm-diameter NbAs nanowire (red dots) and a 40 nm-thick Cu film (black dots). **i,** Room-temperature resistivity as a function of diameter for conventional metal nanowires and NbAs nanowires. The fitted curves for Cu (black), Ru (blue), and Co (green) are from experimental data in literature,[24,48,49] while NbAs nanowires are plotted as red stars.

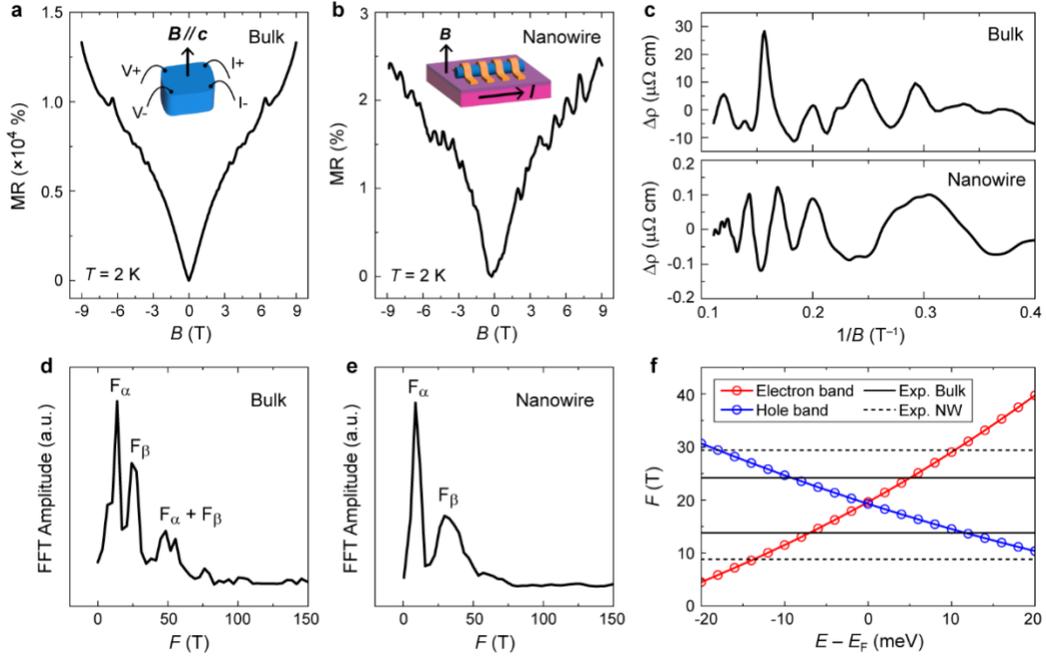

**Fig. 4. Magneto-transport properties of NbAs nanowire and bulk crystal. a, b,** Perpendicular magnetoresistance of a bulk single crystal (**a**) and a nanowire device (**b**) measured at 2 K. **c,** Oscillatory component of the resistivity, obtained by subtracting a polynomial background from the data shown in (**a, b**). **d, e,** FFT of the oscillations for the bulk crystal (**d**) and the nanowire (**e**), with two frequencies ($F_\alpha$ and $F_\beta$) marked. $F_\alpha + F_\beta$ indicates the higher harmonic frequency of $F_\alpha$ and $F_\beta$. **f,** Calculated SdH oscillation frequencies for the electron (red) and hole (blue) pockets of the *bulk* states as a function of the chemical potential. Experimental values are indicated for comparison. The solid black line denotes the SdH oscillation frequencies of the bulk single crystal; the dashed black line denotes those of the nanowire. The SdH oscillation frequencies of the surface states are much higher and exceed the range that our measurement can resolve.

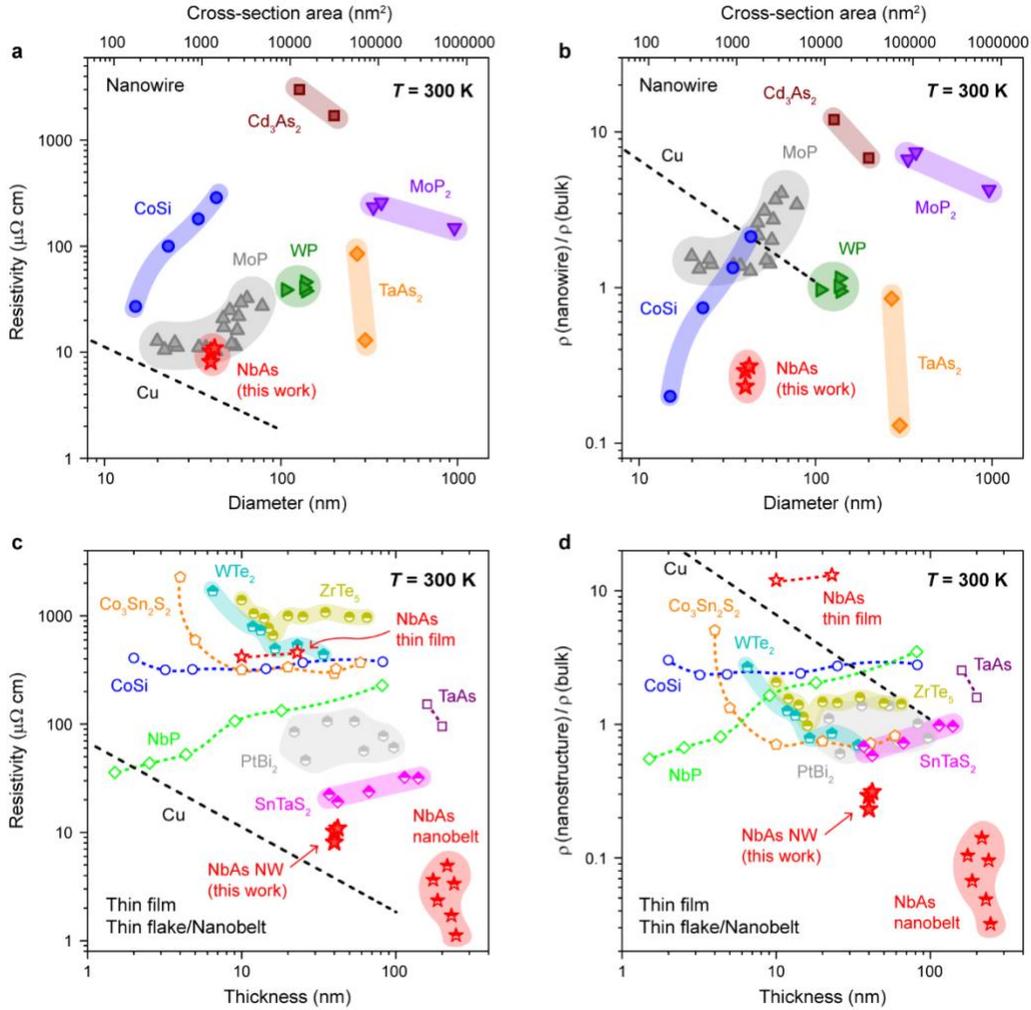

**Fig. 5. Benchmarking NbAs nanowires against other promising topological semimetal nanostructures. a,** Room-temperature electrical resistivity as a function of diameter (cross-section area) for NbAs nanowires and various topological semimetal nanowires (CoSi, MoP, WP, $Cd_3As_2$, $MoP_2$, and $TaAs_2$).[10,43,54-58] For comparison with current interconnect technologies, the resistivity of Cu damascene lines with a Co/TaN liner/barrier layer is also plotted.[24] **b,** Room-temperature resistivities from (**a**), scaled by their respective bulk values, defined as $\rho$(nanowire)/$\rho$(bulk). **c,** Room-temperature electrical resistivity vs. thickness for various topological semimetal thin films and thin flakes/nanobelts. Thin films (NbAs, CoSi, NbP, TaAs, and $Co_3Sn_2S_2$)[12,15,16,59,60] and thin flakes/nanobelts (NbAs, $WTe_2$, $ZrTe_5$, $PtBi_2$, and $SnTaS_2$)[11,61-64] are plotted as open symbols and half-solid symbols, respectively. **d,** Room-temperature resistivities from (**c**), scaled by their respective bulk values, defined as $\rho$(nanostructure)/$\rho$(bulk).